\documentclass[12pt]{article}
\usepackage{amsfonts}
\usepackage{mathrsfs}

\usepackage{multirow}

\usepackage{bbding}
\usepackage{amssymb}
\usepackage{amsmath}
\usepackage{graphicx,color}

\newcommand{\bea}{\begin{eqnarray}}
\newcommand{\eea}{\end{eqnarray}}
\newcommand{\Bea}{\begin{eqnarray*}}
\newcommand{\Eea}{\end{eqnarray*}}
\newcommand{\ba}{\begin{array}}
\newcommand{\ea}{\end{array}}
\newcommand{\bt}{\begin{tabular}}
\newcommand{\et}{\end{tabular}}
\newcommand{\btb}{\begin{table}}
\newcommand{\etb}{\end{table}}
\newcommand{\bc}{\begin{center}}
\newcommand{\ec}{\end{center}}
\newcommand{\beq}{\begin{equation}}
\newcommand{\eeq}{\end{equation}}

\begin{document}

\title{
Estimation and inference
 for  high-dimensional non-sparse models  }
\author{
Lu Lin,  Lixing Zhu\footnote{Lu Lin is a professor of the School of
Mathematics at Shandong University, Jinan, China. His research was
supported by NNSF project (11171188) of China, NBRP (973 Program
2007CB814901) of China, NSF and SRRF projects (ZR2010AZ001 and
BS2011SF006) of Shandong Province of China and K C Wong-HKBU
Fellowship Programme for Mainland China Scholars 2010-11. Lixing Zhu
is a Chair professor of Department of Mathematics at Hong Kong
Baptist University, Hong Kong, China. He was supported by a grant
from the University Grants Council of Hong Kong, Hong Kong, China.
Yujie Gai is an assistant professor of the School of Economics at
Central University of Finance and Economics, Beijing, China. The
first two authors are in charge of the methodology development and
material organization.}\, \,
  and Yujie Gai}
\date{}
\maketitle

\vspace{-3ex}

\begin{abstract}

To successfully work on variable selection, sparse model structure has become a  basic assumption for all existing methods. However, this assumption is questionable as it is hard to hold in most of cases and none of existing methods may provide consistent estimation and
accurate model prediction in nons-parse scenarios. In this paper,   we  propose semiparametric re-modeling
and inference when the linear regression model under study is possibly
non-sparse. After an initial working model is selected by a method such as the Dantzig selector adopted in this paper, we re-construct a
globally unbiased semiparametric model by use of suitable
instrumental variables and nonparametric adjustment. The newly
defined model is identifiable, and the estimator of parameter vector
is asymptotically normal. The consistency, together with the
re-built model, promotes model prediction. This method
naturally works when the model is indeed sparse and thus is of
robustness against non-sparseness in certain sense. Simulation
studies show that the new approach has, particularly when $p$ is
much larger than $n$, significant improvement of estimation and
prediction accuracies over the Gaussian Dantzig selector and other
classical methods. Even when the model under study is sparse, our
method is also comparable to the existing methods designed for
sparse models.

\

{\it Keywords.} Bias correction, 
high-dimensional regression, instrumental variable,  nonparametric
adjustment, re-modeling.

\vspace{1ex}

{\it AMS 2001 subject classification}: 62C05, 62F10, 62F12, 62G05.

\vspace{1ex}

{\it Running head.} Non-sparse models.

\end{abstract}

\baselineskip=20pt

\newpage

\noindent {\large\bf 1. Introduction}

In this paper we consider the linear model
$Y=\beta^{\tau}X+\varepsilon $ as a full model that contains all
possibly relevant predictors $X_1, \cdots, X_p$ in the predictor
vector $X$. Here the dimension $p$ of $X$  is large and even larger
than the sample size $n$. 
As in many cases, most of the predictors are insignificant in a
certain sense for the response $Y$, variable selection is then
necessary. Although this topic has been very intensively
investigated in the literature, the following issues have not yet
received enough attention in the literature.

\begin{itemize}
\item The success of almost all existing variable/feature
section methodologies critically hinges on sparse model structure.
Resulting working model that contains ``significant" predictors is
still assumed to be a linear model having identical model structure
as the full model. Note that this happens only when the full model
has exactly sparse structure. However, in most cases, the full model
may not be exactly sparse. This  then causes that model
identifiability is even an issue. More precisely, after model
selection, resultant working model is usually biased because the
cumulated bias caused by excluding too many ``insignificant"
predictors is non-negligible even when every coefficient associated
with ``insignificant" predictor is indeed very small.  As such, it
is necessary to refine working model so that it becomes unbiased and
identifiable, otherwise the estimator based on it cannot be
consistent and  the prediction would not be accurate. It is worth
pointing out that obviously the refined working model is not
necessary to have identical model structure to  the original full
model unless the full model is sparse. To the best of our knowledge,
there are no research works handling these issues. In this paper, we
will propose a method to reconstruct
 working model, define consistent estimation of the coefficients
associated with the significant predictors contained in the selected model and
further improve prediction accuracy.
\end{itemize}

In this paper, ``non-sparsity" is in the sense that only a few
regression coefficients are large and the rest are small but not
necessary to be zero. A detailed definition on ``non-sparsity" will
be given in the next section for model identification. Furthermore,
it is known that checking either sparsity or non-sparsity of a
high-dimensional model is  a hard task. When there is no prior
information on  sparsity  in advance, as a robustness or
conservative consideration, employing non-sparse model is also
useful for avoiding modelling risk. Of course, when the model under
study is really sparse, it is in hope that  new method also works.

It is  noted that Zhang and Huang (2008) also investigated a model
in which only a few regression coefficients are large and the rest
are small, although they still called it sparse model. In their
paper, the rate consistency was investigated, which means that the
number of selected variables is of the same convergence rate as that
of the variables with large coefficients in an asymptotic sense.
This consistency does not imply the conventional estimation
consistency and does not promote prediction accuracy. This is
because in the scenario they investigated, estimation consistency
and prediction accuracy have not yet been discussed and are still
the challenges. In our paper, by re-modeling selected working model
obtained from the full model, estimation consistency can be achieved
and model prediction accuracy can be improved.

For sparse models there are a great number of research works in the
literature. We list a few here.  The LASSO and the adaptive LASSO
(Tibshirani 1996; Zou 2006),  the SCAD (Fan and Li 2001; Fan and
Peng 2004), the Dantzig selector (Cand\'es and Tao 2007) and MCP
(Zhang 2010) can be used to provide consistent and asymptotically
normally distributed estimation for the
parameters in selected working models. 
In practice, there are no approaches to check sparsity before using them.

To motivate our method, we focus mainly on the Dantzig selector.
The Dantzig selector has received much attention, and 
an asymptotic equivalence between the Dantzig selector and the LASSO
in certain senses was discovered by James {\it et al.} (2009).
Under the uniform uncertainty principle, the resulting estimator
achieves an ideal risk of $\sigma C\sqrt{\log p}$ with a large
probability. This implies that for large $p$, such a risk can be
however large and then even under sparse structure the relevant
estimator may also be inconsistent. To reduce the risk and improve
the performance of relevant estimation, the Gaussian Dantzig
Selector, a two-stage estimation method, was suggested in the literature
(Cand\'es and Tao 2007). The corresponding estimator is still
inconsistent when the model is non-sparse (for details see the next
section). Another method is the Double Dantzig Selector (James and
Radchenko 2009), by which one may choose a more accurate model and,
at the same time, get a more accurate estimator. But  it still
critically depends on the choice of shrinkage tuning parameter and
sparsity condition. Taking these problems into account, Fan and Lv
(2008) introduced a sure independent screening method that is based
on correlation learning to reduce high dimensionality   to a
moderate scale below the sample size. Afterwards,  variable
selection and parameter estimation can be accomplished by
sophisticated method such as the LASSO, the SCAD or the Dantzig
selector. The relevant references include Kosorok and Ma (2007), 
Chen and Qin (2009), James, Radchenko and Lv (2009) and Kuelbs and
Anand (2009), among others. However, when the model is non-sparse
and the dimension $p$ of the predictor vector is very large, the
model is not identifiable and the estimation consistency by existing
methods is usually very difficultly achieved and even not possible.
It causes that model prediction would be less accurate and further
data analysis would
not be reliable unless we can correct bias.

 Thus, for non-sparse model, we have no reasons to expect an unbiased working model that has an identical form to its full model when only a small portion of predictors are regarded as significant and are selected into the working model. Bias correction is necessary.
In this paper, we focus our attention on
working sub-model that is chosen by the Dantzig selector. For the
full model, we will suggest an identifiability condition and a
re-modeling method to identify a working model, and further to
construct consistent and asymptotically normal distributed
estimator for the coefficient vector in the working sub-model. To achieve
this, an adjustment will be recommended to construct a
globally identifiable and unbiased semiparametric model. The
 adjustment only depends on a low-dimensional
nonparametric estimation by using proper instrument variables. The
resulting estimator $\hat\theta$ of the parameter vector $\theta$ in
the sub-model satisfies
$\|\hat\theta-\theta\|^2_{{\ell}_2}=O_p(n^{-1})$ and the asymptotic
normality if the dimension $q$ of $\theta$ converges to a fixed constant with a probability tending to one. Furthermore,
new consistent estimators  together with the unbiased adjustment
sub-model or the original sub-model defined in this paper, can also improve model
prediction accuracy.  This is the first attempt in this area for us to understand modeling after variable selection when sparse structure is not imposed. It is worth mentioning that although  insignificant predictors are ruled out in the selection step, we do not absolutely abandon them, while use them to construct adjustment variables.

It is worth pointing out that the newly proposed method is a general
method which may also be applicable with other variable selection
approaches. On the other hand, the new method is robust against non-sparseness at the
cost that the new algorithm is slightly more complicated to implement
than existing methods are because we transfer a linear model to a nonlinear model.  However to
avoid the risk of possible unreliable further analysis caused by the inconsistency of estimation and promote more accurate prediction, such a cost
is worthwhile to pay.

The rest of the paper is organized as follows. In Section~2  the
properties of the Dantzig estimator for the high-dimensional linear
model are reviewed. In Section~3, an identifiability condition is
assumed, a bias-corrected sub-model is proposed via introducing
instrumental variables, and a nonparametric adjustment and a method
about selecting instrumental variables are suggested. Estimation and
prediction procedures for the new sub-model are given and the
asymptotic properties of the resulting estimator and prediction are
obtained. In Section~4 an approximate algorithm for constructing
instrumental variables is proposed for the case when the dimension
of the related nonparametric estimation is relatively large.
Simulation studies are presented in Section~5 to examine the
performance of the new approach when compared with the classical
Dantzig selector and other methods. The technical proofs for the
theoretical results are provided in the online supplement to this
article.

\

\noindent {\large\bf 2. A brief review for the Dantzig selector}

Recall the full model:
$$Y=\beta^{\tau}X+\varepsilon,\eqno(2.1)$$ where $Y$ is the scale
response, $X$ is the $p$-dimensional predictor and $\varepsilon$ is
the random error satisfying $E(\varepsilon|X)=0$ and
$Cov(\varepsilon|X)=\sigma^2$. 
Throughout this paper, of the
primary interest is to build a valid sub-model of (2.1) whose size goes to a
non-random number with a probability tending to one. Non-randomness of selected sub-model is for further model identifiability. We then  build an
adjusted model that is unbiased and identifiable. The second
interest of our paper is to construct consistent estimators for
significant predictors in the rebuilt model and further to obtain
reasonable model prediction via our estimation and selected
sub-model or adjusted model.

To introduce a re-modeling method and a novel estimation approach, we first
re-examine the Dantzig selector. Let ${\bf
Y}=(Y_1,\cdots,Y_n)^{\tau}$ be the vector of the observed responses
and ${\bf X}=(X_1,\cdots,X_n)^{\tau}=({\bf x}_1,\cdots,{\bf x}_p)$
be the
 $n\times p$ matrix of the observed predictors. The Dantzig selector of $\beta$ is defined as
$$\tilde \beta^D=\arg\min_{\beta\in\mathscr{B}}\|\beta\|_{{\ell}_1} \ \
\mbox{ subject to } \ \ \sup_{1\leq j\leq p}|{\bf x}^{\tau}_jr| \leq
\lambda_p\,\sigma\eqno(2.2)$$ for some $\lambda_p>0$, where
$\|\beta\|_{{\ell}_1}=\sum_{j=1}^p|\beta_j|$ and $r={\bf Y}-{\bf
X}\beta$. As was shown by Cand\'es and Tao (2007), under sparsity assumption and other regularity
conditions, this estimator satisfies that, with large probability,
$$\|\tilde\beta^D-\beta\|^2_{{\ell}_2}\leq  C  \sigma^2\log p,\eqno(2.3)$$
where $C$ is free of $p$ and $\|\tilde\beta^D-\beta\|^2_{{\ell}_2}=
\sum_{j=1}^p(\tilde \beta_j^D-\beta_j)^2$. In fact this is an ideal
risk and thus cannot be improved in a certain sense. However,  such
a risk can become large and may not be negligible when the dimension
$p>n$. On the other hand, if without sparsity condition, the risk
will be even larger than that given in (2.3).

To reduce the risk and promote the performance of the Dantzig selector, one often uses a two-stage selection procedure (e. g., the
Gaussian Dantzig Selector) to construct a risk-reduced estimator for
the obtained sub-model (Cand\'es and Tao 2007). For example, we can
first estimate $I=\{j:\beta_j\neq 0\}$ with $\tilde I=\{j:|\tilde
\beta_j^D|>\kappa\sigma\}$ for some $\kappa\geq 0$ and then
construct an estimator
$$\tilde\beta_{\tilde I}=(({\bf X}^{(\tilde I)})^{\tau}{\bf X}^{(\tilde I)})^{-1}({\bf X}^{(\tilde
I)})^{\tau}{\bf Y}$$ for $\beta_{ I}$ and shrink the other components
of $\beta$ to be zero, where $\beta_{\tilde I}$ is the restriction
of $\beta$ to the set ${\tilde I}$, and ${\bf X}^{(\tilde I)}$ is
the matrix with the column vectors according to $\tilde I$.

When model is not sparse, the set $I$ is very large and there is no method available in the literature to consistently estimate $\beta_I$.  However, for variable / feature selection, we  are mainly interested in those significant variables that are associated with large values of coefficients. Thus,
denote $\beta_{\tilde I}=\theta$, a $q$-dimensional vector of
interest.  To identify the set $I$, we will give an identifiability condition to ensure that the random set $\tilde I$ converges to $I$ with probability tending to one. For the sake of description, we temporarily assume that $\tilde I$ is fixed.  Without loss of generality, suppose that $\beta$ can be
partitioned as $\beta=(\theta^{\tau},\gamma^{\tau})^{\tau}$ and, correspondingly, $X$
is partitioned as $X=(Z^{\tau},U^{\tau})^{\tau}$. Then the above two-stage procedure
implies that based on the Dantzig selector, we  use the sub-model
$$Y=\theta^{\tau}Z+\eta\eqno(2.4)$$ to replace the full-model (2.1), where
$\eta=\gamma^{\tau}U+\varepsilon$ is regarded as error. Here the dimension
$q$ of $\theta$ can be either fixed or diverging with $n$ at a
certain rate. Since the above sub-model is a replacer of the full
model (2.1), we call $\theta$ and $Z$ the main parts of $\beta$ and
$X$, respectively. From (2.1) and (2.4) it follows that $E(\eta|Z)=
\gamma^{\tau}E(U|Z)$. When both $\gamma \neq 0$ and $E(U|Z)\neq 0$, the
sub-model (2.4) is biased and thus the two-stage estimator
$\tilde\theta_S=\tilde\beta_{\tilde I}$ is also biased. It shows
that the two-stage estimator $\tilde\theta_S$ of $\theta$ is also
inconsistent. Note that for any non-sparse model,
$\gamma \neq 0$ always holds. As such, the above classical  method is not possible
to  obtain consistent estimation.

An improved Dantzig selector is the Double
Dantzig Selector (James and
Radchenko 2009). By which more accurate model and estimation can be
expected. In the first step, the Dantzig selector is used with a
relatively large shrinkage tuning parameter $\lambda_p$ defined
above to get a relatively accurate sub-model in the sense that less
insignificant predictors are contained. The Dantzig selector is
further used in the selected sub-model to obtain a relatively
accurate estimator of $\theta$ via a small $\lambda_p$ and data
$(Y,Z)$.  However, such a method cannot handle non-sparse model either
because the sub-model selected in the first step has already been
biased. It is also noted that this method critically depends on
twice choices of shrinkage tuning parameter $\lambda_p$; for details
see James and Radchenko (2009). On the other hand, when the estimation
consistency and asymptotic normality, rather than variable selection,  heavily
depend on the choice of $\lambda_p$, it is practically not
convenient, and more seriously, the consistency is in effect not
judgeable unless a criterion of tuning parameter selection can be
defined
to ensure the consistency.  %

\noindent {\large\bf 3. Re-modeling and inference}

As was shown above, the sub-model (2.4) is usually biased and random
after the variable selection determined by the Dantzig selector. Here the
model randomicity means that the estimate $\tilde I$ for the index
set $I$ defined in the previous section is random. As this section is long containing the main contributions, we separate it into several subsections. we first propose an identifiability condition for non-sparse models; subsection~3.2 investigates a re-modeling scheme; the estimation procedure is described in subsection~3.3. To highlight the procedure, we have a short subsection~3.4 to summarize the steps of the algorithm. The asymptotic behaviours are put in subsection~3.5. Subsection~3.6 discusses the prediction issue.

\noindent{\bf 3.1 Identifiability condition.} Before re-modeling
and inference, we first assume a condition to guarantee that the
working sub-model (2.4) is identifiable  with probability
approaching one. Let $|J|$ be the number of elements in an index set
$J\subset \{1,2,\cdots,p\}$ and $\bar J$ be the complement of $J$ in
the set $\{1,2,\cdots,p\}$. For a $p$-dimensional vector
$l=(l_1,\cdots,l_p)^{\tau}$, denote by $l_J=(l_j)_{j\in J}$ a
subvector whose entries are those of $l$ indexed by
 $J$.

\begin{itemize}
\item[(C0)] {\it Identifiability condition:}
\begin{itemize} \item[1)] Index set $I$ satisfies that $\min_{j\in I}|\beta_j|\geq c n^{(c_1-1)/2}$
and
$$ \min_{l_I\neq 0, \|l_{\bar I}\|_{\ell_1}\leq
\|l_I\|_{\ell_1}+2c_2n^{(c_1-1)/2}}\frac{\|Xl\|_{{\ell}_2}}{\sqrt{n}\|l_I\|_{{\ell}_2}}>\sqrt{{3}/{8}},
\ \ (a.s.) \eqno(3.1)
$$
where constants $0<c_1<1$, $c_2>0$, $c=4kbq\sigma+4\sqrt{k^2b^2q^2
\sigma^2+3kc_2bq\sigma/8}$, $b>\sqrt{2}$, $q=|I|$ and $k>0$.
\item[2)]  $\bar I$
satisfies that $\|\beta_{\bar I}\|_{\ell_1}=c_2n^{(c_1-1)/2}$
 and $\max_{j\in \bar I}|\beta_j|=o(n^{(c_1-1)/2})$.
 \end{itemize}
\end{itemize}

Part 1) of condition (C0) means that the coefficients
in the selected set $I$ are significant and the inequality (3.1) is
to control the restricted eigenvalues. Such an inequality is similar
to the assumption in Bickel {\it et al}. (2009). Part 2) means the
non-sparsity in the following sense: the coefficients that are
associated with insignificant predictors  may not be exactly zero but
decays to zero at the rate of $n^{(c_1-1)/2}$ as the sample size $n$
goes to infinity. We can easily construct non-sparse models
satisfying condition (C0). Under this non-sparse condition, all significant regression
coefficients are contained in the selected set $I$ in an asymptotic
sense and therefore model identifiability is achieved when we select
a working sub-model; for details see the following model selection
principle and lemma.

With condition (C0), we could select a set of indices as
$$\tilde I_{\tau_n}=\{1\leq j\leq p:|\tilde\beta_j^D|\geq
\tau_n\},$$ where $\tau_n$ is a predefined threshold value so that
the obtained sub-model (2.4) is non-random with probability
approaching one; the following lemma presents the details.

{\bf Lemma 3.1} {\it In addition to Condition (C0), assume that  $\sqrt{\log
p}=kn^{c_1/2}$ with
$\frac{72c_2}{51bq\sigma}<k<\min\{\frac{9(c_3-2c_2)^4}{32\times
64(c_3-c_2)^2\sigma^2q^2},\frac{3c_2}{2qb\sigma}\}$ and $c_3>2c_2$,
all the diagonal elements of the matrix ${\bf X}^{\tau}{\bf X}/n$
are equal to 1, $\lambda=b\sigma\sqrt{\frac{\log p}{n}}$ and
$\tau_n=\frac{c}{2} n^{(c_1-1)/2}$. Then as $n \rightarrow \infty$
$$P(\tilde I_{\tau_n}=I)\rightarrow 1.$$
}

The proof of the lemma is given in the Appendix. We use the
condition on ${\bf X}^{\tau}{\bf X}$ only for the simplicity of
proof. This lemma guarantees that, even the full model is
non-sparse, the selected model equals the  model with all
significant predictors with  probability tending to 1, i.e, the
model selection is asymptotically exact and, therefore, the
sub-model (2.4) could be regarded as a non-random model.

\noindent{\bf 3.2 Re-modeling.} It is obvious that remodeling for bias correction
is necessary to the selected sub-model (2.4) when we
want to get a valid model and have  consistent estimation for the
sub-vector $\theta=(\theta_1,\cdots,\theta_q)^{\tau}$.  To this end,
a new model with an instrumental variable is established in this
subsection. Suppose that the $q$ significant predictors can be selected with probability going to one, which  will be proved later. Denote
$Z^\star=(Z^{\tau},U^{(1)},\cdots,U^{(d)})^{\tau}$ and $W=AZ^\star$,
where $A$ is $r\times (q+d)$ matrix satisfying that its row vectors
have length 1. Here $U^{(1)},\cdots,U^{(d)}$ are pseudo-variables
(or instrumental variables), and, without loss of generality, they
are supposed to be the first $d$ components of $U$. It will be seen
that we choose $d=1$ usually. Set
$V=(\alpha^{\tau}U,W^{\tau})^{\tau}$, where $\alpha$ is a vector to
be chosen later. Choose $A$ and $U^{(1)},\cdots,U^{(d)}$ such that
$$E\{(Z-E(Z|V))(Z-E(Z|V))^{\tau}\}>0.\eqno(3.2)$$
This condition on the matrix we need can trivially hold because $V$
contains $W$ that is a weighted sum of $Z$ and
$U^{(1)},\cdots,U^{(d)}$. The use of condition (3.2) is to guarantee
the identifiability of the following model. The choice of $\alpha$, $A$ and $U$ will be discussed later.

Denote $g(V)=E(\eta|V)$.  Now we introduce a
bias-corrected version of (2.4) as
$$Y_i=\theta^{\tau}Z_i+g(V_i)+\xi(V_i),\ i=1,\cdots,n, \eqno(3.3)$$ where $\xi(V)
=\eta-g(V)$. Obviously, if $\alpha$ in $V$ is identical to $\gamma$
in $\eta$, this model is unbiased, i.e., $E(\xi|Z,V)=0$; otherwise
it may be biased. This model can be regarded as a partially linear
model with a linear component $\theta^{\tau}Z$ and a nonparametric
component $g(V)$, and is identifiable because of condition
(3.2). From this structure, we can see that when $V$ does not
contain the instrumental variable $W$ and $\alpha=\gamma$, the model
goes back to the original working model of (2.4) as $\xi $ is zero and $g(V)$
becomes the error term $\eta$ (if $\varepsilon$ is ignored). This
observation motivates us to consider the following method. Introducing an instrumental
variable  $V$ so that $\xi$ has a zero conditional mean, we
can estimate $g(\cdot)$ so that we can correct the bias occurred in the original working
model. Although a nonparametric function $g(v)$ is involved, it will
be verified that the dimension $r+1$ of the variable $v$ may be low
usually. For the case of large $r$, we will introduce an approximate
method to deal with the problem. Note that for $V$, the key is to
properly select $\alpha$ and $W$. From the above description, we can
see that although $\alpha=\gamma$ should be a natural and good
choice, it is unknown and cannot be estimated consistently when the
dimension is large. Taking this into account, we first consider a
general $\alpha$ and construct a bias-corrected model with suitable
$W$, or equivalently a suitable matrix $A$.

To this end, we need the condition that $(Z,U)$ is elliptically symmetrically distributed. The ellipticity condition can be slightly weakened to be the following linearity condition:
$$E(U|C^{\tau}Z^\star)=E(U)+
\Sigma_{U,Z^\star}C(C^{\tau}\Sigma_{Z^\star,Z^\star}C)^{-1}C^{\tau}(Z^\star-E(Z^\star))$$ for some given matrix $C$.
The linearity
condition has been widely assumed in the circumstance of high-dimensional
models. Hall and Li (1993) showed that it
often holds approximately when the dimension $p$ is high.

With the above condition, we can find a matrix $A$ so that the model
(3.3) is always unbiased. Let
$\Sigma_{Z^\star,Z^\star}=Cov(Z^\star,Z^\star)$ and
$\Sigma_{U,Z^\star} =Cov(U,Z^\star)$. Denote by $r$ the rank of
matrix $\Sigma_{U,Z^\star}$. Obviously, $r$ is bounded if $q$ is
fixed because in this case the dimension of matrix $Z^\star$ is
bounded. It is known by singular value decomposition of matrix that
$$\Sigma_{U,Z^\star}=P\left(\begin{array}{ll}\Lambda_r&0\\0&0\end{array}\right)Q^\tau,$$
where $P$ is a $(p-d)\times(p-d)$ orthogonal matrix, $Q$ is a
$d\times d$ orthogonal matrix and
$\Lambda_r=\mbox{diag}(\eta_1,\cdots,\eta_r)$ with $\eta_j>0$ and
$\eta_j^2$ being positive eigenvalues of $\Sigma_{U,Z^\star}^\tau
\Sigma_{U,Z^\star}$. Let $Q=(Q_1,Q_2)$, where $Q_1$ is a $d\times r$
orthogonal matrix. In this case, we have the following conclusion.

{\bf Lemma 3.2} {\it Under the above linearity condition, when
$\Sigma_{Z^\star,Z^\star}=I_{q+d}$ and
$$A=Q_1^{\tau},\eqno(3.4)$$  the model (3.3) is then unbiased, that is, $E(\xi|Z,V)=0$.}

The condition $\Sigma_{Z^\star,Z^\star}=I_{q+d}$  is
common because the components of $Z^\star$ that are selected from $X$ form a low-dimensional matrix. The proof of the lemma is
presented in Appendix. This lemma ensures that, with such a choice
of $A$, the model (3.3) is always unbiased whether the model (2.1)
is sparse or not.

The covariance matrix $\Sigma_{U,Z^\star}$ is not always given and
then needs to be estimated. It is known that the methods for
constructing consistent estimation for large covariance matrix have
been proposed in the  literature, for example the tapering
estimators investigated by Cai, Zhang and Zhou (2010). Let
$\hat\Sigma_{U,Z^\star}$ be a consistent estimator of
$\Sigma_{U,Z^\star}$, satisfying
$$\|\hat\Sigma_{U,Z^\star}-\Sigma_{U,Z^\star}\|=O_p(n^{-\varsigma}),\eqno(3.5)$$
where constant $\varsigma>0$ and $\|\cdot\|$ is a matrix norm. By
the singular value decomposition of matrix mentioned above, we get
an estimator of $Q_1$ as $\hat Q_1$. Then $\hat A=\hat Q_1^\tau$ is
a consistent estimator of $A$, satisfying
$$\|\hat A-A\|=O_p(n^{-\varsigma}).$$

From the above choice of $A$, we can see that $g(v)$ is a
$(r+1)$-variate nonparametric function. To realize the estimation
procedure and reduce the dimension of variable $v$, we choose a
threshold $\upsilon_n>0$ and then set $\hat\phi_j=0$ if
$\hat\phi_j<\upsilon_n$. Suppose that $\hat\phi_1\geq\cdots\geq
\hat\phi_{r^*}\geq\upsilon_n$ and the corresponding orthogonal
matrix is $\hat Q^*_1$,  where $r^*\leq r$ and $\hat Q^*_1$ is a
$(q+d)\times r^*$ matrix. In this case, the estimator of $A$ is
$\hat A={\hat {Q^*}_1}'$ and as a result, $g(v)$ is a
$(r^*+1)$-variate nonparametric function, in which the dimension of
the variable is lower than or equal to the original one. Usually we
choose $d=1$, and similar to Irrepresentable condition (Zhao and Yu
2006), we may assume that the rank of covariance matrix of $(Z,U)$ is low
(equivalently, the correlation between $Z$ and $U$ is weak). In this
case $g(v)$ can be a low-dimensional nonparametric function. If $r^*$ is
still large, we use a row vector to replace $A$ and will give a
method in Section~4 to find an approximate solution with which
$g(v)$ is a 2-dimensional nonparametric function.

The above deduction and justification show that the above bias-correction procedure
is free of the choice of $\alpha$. However, choosing a proper
$\alpha$ is of importance. An ideal choice of $\alpha$ should be as
close to $\gamma$ as possible. In the estimation procedure, a
natural choice is the estimator $\tilde \gamma^D$ of $\gamma$, which
is obtained in the step of using the Dantzig selector. Also we will discuss the
asymptotic properties of the estimator of $\theta$ for both the  cases
where $\alpha$ is  given and is estimated respectively in Subsection~3.4.

\noindent{\bf 3.3 Estimation.} Recall that the bias-corrected model
(3.3) can be thought of as a partially linear model. We therefore
design an estimation procedure as follows. First of all, as
mentioned above, for any $\alpha$, the model (3.3) is unbiased. Then
we can design the estimation procedure when $\alpha$ has been determined
by any empirical method. Given $\theta$ and for any $\alpha$, if $A$
is estimated by $\hat A$, then the nonparametric function $g(v)$ is
estimated by
$$\hat g_\theta(v)=\frac{\sum_{k=1}^n(Y_k-\theta^{\tau}
Z_k)L_H(\hat V_k-v)} {\sum_{k=1}^nL_H(\hat V_k-v)},$$ where $\hat
V=(\alpha^{\tau}U,\hat W^{\tau})^{\tau}$ with $\hat W=\hat
AZ^\star$, $L_H(\cdot)$ is a $(r+1)$-dimensional kernel function. A
simple choice of $L_H(\cdot)$ is a product kernel as
$$L_H(V-v)=\frac{1}{h^{r+1}}K\Big(\frac{V^{(1)}-v^{(1)}}{h}\Big)\cdots
K\Big(\frac{V^{(r+1)}-v^{(r+1)}}{h}\Big),$$ where
$V^{(j)},j=1,\cdots,r+1$, are the components of $V$, $K(\cdot)$ is
an 1-dimensional kernel function and $h$ is the bandwidth depending
on $n$. Particularly, when $\alpha$ is chosen as $\tilde \gamma^D$,
we get an  estimator of $g(v)$ as
$$\tilde g_\theta(v)=\frac{\sum_{k=1}^n(Y_k-\theta^{\tau} Z_k)L_H(\tilde V_k-v)}
{\sum_{k=1}^nL_H(\tilde V_k-v)},$$ where $\tilde
V=(U^{\tau}\tilde\gamma^D,\hat W^{\tau})^{\tau}$.

With the two estimators of $g(v)$, the
bias-corrected model (3.3) can be approximately expressed by the
following two models:
$$Y_i\approx \theta^{\tau}Z_i+\hat g_\theta(\hat V_i)+\xi(\hat V_i)\ \ \mbox{ and }\ \
Y_i\approx \theta^{\tau}Z_i+\tilde g_\theta(\tilde V_i)+\xi(\tilde
V_i),$$ equivalently,
$$\hat Y_i\approx \theta^{\tau}\hat Z_i+\xi(\hat V_i)\ \ \mbox{ and }\ \ \tilde Y_i\approx \theta^{\tau}\tilde Z_i+\xi(\tilde V_i),\eqno(3.6)$$
where
$$\hat Y_i=Y_i-\frac{\sum_{k=1}^nY_kL_H(\hat V_k-\hat V_i)}
{\sum_{k=1}^nL_H( \hat V_k-\hat V_i)},\ \ \hat
Z_i=Z_i-\frac{\sum_{k=1}^nZ_kL_H(\hat V_k-\hat
V_i)}{\sum_{k=1}^nL_H( \hat V_k- \hat V_i)},$$
$$\tilde Y_i=Y_i-\frac{\sum_{k=1}^nY_kL_H(\tilde V_k-\tilde V_i)}
{\sum_{k=1}^nL_H(\tilde V_k-\tilde V_i)},\ \ \tilde
Z_i=Z_i-\frac{\sum_{k=1}^nZ_kL_H(\tilde V_k-\tilde
V_i)}{\sum_{k=1}^nL_H(\tilde V_k-\tilde V_i)}.$$ Thus,  the
sub-models in (3.6) result in two estimators of $\theta$ as
$$ \hat\theta=\hat S_n^{-1}\frac{1}{n}\sum_{i=1}^n \hat
Z_i\hat Y_i \ \ \mbox{ and }\ \ \tilde\theta=\tilde
S_n^{-1}\frac{1}{n}\sum_{i=1}^n \tilde Z_i\tilde Y_i,\eqno(3.7)$$
where $\hat S_n=\frac{1}{n}\sum_{i=1}^n\hat Z_i\hat Z_i^{\tau}$ and
$\tilde S_n=\frac{1}{n}\sum_{i=1}^n\tilde Z_i\tilde Z_i^{\tau}$,
respectively. Here we assume that the bias-corrected model (3.3) is
homoscedastic, that is $Var(\xi(\hat V_i))=\sigma_V^2$ and
$Var(\tilde \xi(V_i))=\sigma_V^2$ for all $i=1,\cdots,n$. If the
model is heteroscedastic, we respectively modify the above
estimators as,assuming that  $\sigma_i^2(\hat V_i)$ and $\sigma_i^2(\tilde V_i)$ are known,
$$\hat\theta^*=\hat {S_n^*}^{-1}\frac{1}{n}\sum_{i=1}^n
\frac{1}{\sigma_i^2(\hat V_i)}\hat Z_i\hat Y_i
 \ \ \mbox{ and }\ \ \tilde\theta^*=\tilde{S_n^*}^{-1}\frac{1}{n}\sum_{i=1}^n
\frac{1}{\sigma_i^2(\tilde V_i)}\tilde Z_i\tilde Y_i,$$ where $\hat
S_n^*=\frac{1}{n}\sum_{i=1}^n\frac{1}{\sigma_i^2(\hat V_i)}\hat
Z_i\hat Z_i^{\tau}$ and $\tilde
S_n^*=\frac{1}{n}\sum_{i=1}^n\frac{1}{\sigma_i^2(\tilde V_i)}\tilde
Z_i\tilde Z_i^{\tau}$, respectively, and $\sigma_i^2(\hat
V_i)=Var(\xi(\hat V_i))$ and $\sigma_i^2(\tilde  V_i)=Var(\xi(\tilde
V_i))$.  When $\sigma_i^2(\hat V_i)$ and $\sigma_i^2(\tilde V_i)$ are unknown, we can use their
consistent estimators to replace them; for details about how to
estimate them see for example H\"ardle {\it et al.} (2000). In the
following we only consider the estimators defined in (3.7). Finally,
an estimator of $g(v)$ can be defined as either $\hat
g_{\hat\theta}(v)$ or $\tilde g_{\tilde\theta}(v)$.

\noindent{\bf 3.4 Algorithm.}
In summary, our algorithm procedure includes following three steps:

{\bf Step 1}. Choose an initial value of $\alpha$, which may be arbitrary or
estimated.

{\bf Step 2}. Decompose matrix $\Sigma_{U,Z^\star}$ (singular value
decomposition) and then choose $A=Q_1^{\tau}$ or $A=\hat
Q_1^{\tau}$, an estimator of $Q_1^{\tau}$, if $\Sigma_{U,Z^\star}$
is unknown.

{\bf Step 3}. Construct estimators by (3.7).

The procedure shows that the new algorithm is slightly more complicated
to implement than existing methods are by transferring an estimation procedure for linear model to that for nonlinear model. However, such a way can obtain  consistent estimation and  promote prediction accuracy for non-sparse model, and thus it is worthwhile to pay the expenses of computation.

\noindent{\bf 3.5 Asymptotic normality.} To study this asymptotic behavior, the following conditions for the
model (3.3)  are assumed:

\begin{itemize}
\item[(C1)] The first two derivatives of $g(v)$ and $\xi(v)$ are continuous.
\item[(C2)] Kernel function
$K(\cdot)$ satisfies $$\int K(u)du=1, \int
u^jK(u)du=0,j=1,\cdots,k-1,0< \int u^kK(u)du<\infty.$$
\item[(C3)] The bandwidth $h$ is optimally chosen, i.e., $h=O(n^{-1/(2k+r+1)})$.
\item[(C4)] The constant $\varsigma$ in (3.5) satisfies $\varsigma>
1/4$.
\end{itemize}

Obviously, conditions (C1)-(C3) are commonly used for
semiparametric models. Condition (C4) is also satisfied for the
consistency of covariance estimators, for example the tapering
estimators investigated by Cai, Zhang and Zhou (2010). With these
conditions, the following theorem states the asymptotic normality
for the bias-corrected estimator $\hat\theta$.

{\bf Theorem 3.3} {\it In addition to the conditions in lemma~3.1, assume that conditions (C1)-(C4) and
(3.2) hold. For a given nonzero vector $\alpha$, if $q$ is fixed and
$p$ may be larger than $n$, then, as $n\rightarrow\infty$,
$$\sqrt{n}(\hat\theta-\theta)\stackrel{D}\longrightarrow N(0,\sigma_V^2S^{-1}),$$
where $S=E\{(Z-E(Z|V))(Z-E(Z|V))^{\tau}\}$.

}

The proof for the theorem is postponed to the Appendix.

{\bf Remark~3.1.} {\it This theorem shows that the new  estimator
$\hat\theta$ is $\sqrt{n}$-consistent regardless of the choice of
the shrinkage tuning parameter $\lambda_p$ and thus it is convenient
to be used in practice. Furthermore, by the theorem and the commonly
used nonparametric techniques, we can prove that $\hat
g_{\hat\theta}(v)$ is also consistent. In effect, we can obtain the
strong consistency and the consistency of the mean squared error
under some stronger conditions. The details are omitted in this
paper. Note that these results can obviously  hold when the model is sparse. Thus, for either sparse or non-sparse model, our method always ensures the estimation consistency for coefficients selected into the working model. }


To investigate the asymptotic properties for the second estimator
$\tilde\theta$ in (3.7) that is based on the Dantzig selector
$\tilde\gamma^D$, we need the following more condition:

\begin{itemize}
\item[(C5)] The maximum eigenvalue $\lambda_M$ of $UU'$ is bounded for all $n$.
\item[(C6)] Suppose that there exists a nonzero vector, say $\alpha$, such that
$\|\tilde\gamma^D-\alpha\|_{{\ell}_2}=O_p(n^{-\mu})$ for some $\mu$
satisfying $\mu>1/4.$
\end{itemize}

Condition (C5) is commonly used for high dimensional models (see,
e.g., Fan and Peng 2004). For condition (C6), we have the following
explanations. As was stated in the previous sections, we use
$\alpha$ to denote an arbitrary vector. The vector $\alpha$ in
condition (C6) is then different from that used before; here
$\alpha$ is a fixed vector. For the simplicity of representation we
still use the same notation $\alpha$ in different appearances.
Condition (C6) is the key for the following theorem. This condition
does not mean that the Dantzig selector $\tilde\gamma^D$ is
consistent. The condition implies that when $n$ is large enough,
$\tilde\gamma^D$ is close to a non-random vector $\alpha$
asymptotically. Note that the accuracy of the solution of linear
programming can guarantee that
$\|\tilde\gamma^D-\alpha\|_{{\ell}_2}$ is small enough for a
solution of the linear programming problem of  (2.2) (see for
example Malgouyres and Zeng, 2009). These show that condition (C6)
is reasonable. Condition (C6) can actually be weakened, but for the
simplicity of technical proof and presentation, we still use the
current conditions in this paper.

{\bf Theorem~3.4} {\it Under conditions (C1)-(C6) and  the conditions in Lemma~3.1,, when $q$ is
fixed and $p$ may be larger than $n$,  we have
$$\sqrt{n}(\tilde\theta-\theta)\stackrel{D}\longrightarrow N(0,\sigma_V^2S^{-1}).$$
}

The proof of the theorem is given in the Appendix.

{\bf Remark~3.2.} {\it This theorem shows that when $\gamma$ is replaced by
the Dantzig selector $\tilde \gamma^D$, the resulting estimator
$\tilde\theta$ is also $\sqrt{n}$-consistent regardless of the choice of
the shrinkage tuning parameter $\lambda_p$. On the other hand,
although Theorems~3.3 and  3.4 have an identical representation for
the asymptotic covariances, the asymptotic covariances of the two
estimators are in fact different because $\alpha$ and therefore $V$
used in the two theorems are different.}


\noindent{\bf 3.6 Prediction.} Combining the estimation consistency
with the unbiasedness of the adjusted sub-model (3.3), we obtain an
improved prediction as
$$\hat
Y=\hat\theta^{\tau}Z+\hat g_{\hat\theta}(V)\eqno(3.8)$$ and the
corresponding prediction error is
$$\begin{array}{lll}E(Y-\hat Y)^2&=&E((\hat\theta-\theta)^{\tau}Z)^2+E(\hat
g_{\hat\theta}(V)-g(V))^2+E(\xi^2(
V))\vspace{1ex}\\&&+2E((\hat\theta-\theta)^{\tau}Z(\hat
g_{\hat\theta}(V)-g(V)))+2E((\hat\theta-\theta)^{\tau}Z\xi(
V))\vspace{1ex}\\&&+2E((\hat g_{\hat\theta}(V)-g(V))\xi(
V))\vspace{1ex}\\&=&E(\xi^2( V))+o(1).\end{array}$$ It is of a
smaller prediction error than the one obtained by the classical Dantzig
selector, and interestingly  any high-dimensional
nonparametric estimation is not needed.

In contrast, the resulting  prediction is defined as, when we use the new estimator $\hat\theta$ and the sub-model (2.4),
rather than the adjusted sub-model (3.3),
$$\hat Y_S=\hat\theta^{\tau}Z+\bar{\hat g}_{\hat\theta},\eqno(3.9)$$
where $$\bar{\hat g}_{\hat\theta}=\frac{1}{n}\sum_{i=1}^n\hat
g_{\hat\theta}(V_i).$$ We  add $\bar{\hat
g}_{\hat\theta}$ in (3.9) for prediction because the sub-model (2.4) has a bias
$E(g(V))$, otherwise, the prediction error would be even larger.  In
this case, $\bar{\hat g}_{\hat\theta}$ is free of the predictor $U$
and the resultant prediction of (3.9) only uses the predictor $Z$ in
the sub-model (2.4). This  is different from the prediction (3.8)
that depends on both the low-dimensional predictor $Z$ and
high-dimensional predictor $U$. Thus (3.9)
is a sub-model based prediction. 
The
corresponding prediction error is
$$\begin{array}{lll}E(Y-\hat
Y_S)^2&=&E((\hat\theta-\theta)^{\tau}Z)^2+E(\bar{\hat
g}_{\hat\theta}-g(V))^2+E(\xi^2(
V))\vspace{1ex}\\&&+2E((\hat\theta-\theta)^{\tau}Z(\bar{\hat
g}_{\hat\theta}-g(V)))+2E((\hat\theta-\theta)^{\tau}Z\xi(
V))\vspace{1ex}\\&&+2E((\bar{\hat g}_{\hat\theta}-g(V))\xi(
V))\vspace{1ex}\\&=&E(\xi^2( V))+Var(g(V))+2E(E(g(V))-g(V))\xi(
V))+o(1).\end{array}$$ This error is usually  larger than that of
the prediction (3.8). However, we can see that
$$|E(E(g(V))-g(V))\xi(
V))|\leq (Var(g(V))Var(\xi(V)))^{1/2}$$ and usually the values of
both $Var(g(V))$ and $Var(\xi(V))$ are small. Then such a prediction
still has a smaller prediction error than the one obtained by the
sub-model (2.4) and the common LS estimator $\tilde\theta_S=({\bf
Z}^{\tau}{\bf Z})^{-1}{\bf Z}^{\tau}{\bf Y}$  as: $$\tilde
Y_S=\tilde\theta_S^{\tau}Z.\eqno(3.10)$$ Precisely,  the corresponding error of $\tilde
Y_S$ in (3.10) is
$$\begin{array}{lll}E(Y-\tilde Y_S)^2&=&E((\tilde\theta_S-\theta)^{\tau}Z)^2+E(\gamma^{\tau}U)^2+\sigma^2
+2E((\tilde\theta_S-\theta)^{\tau}Z\gamma^{\tau}U).\end{array}$$ Because
$\tilde\theta_S$ does not converge to $\theta$, the values of both
$E((\tilde\theta_S-\theta)^{\tau}Z)^2$ and
$2E((\tilde\theta_S-\theta)^{\tau}Z\gamma^{\tau}U)$ are large and as a result
the prediction error is large as well.

The above results show that in the scope of prediction, the new
estimator can reduce prediction error under both the adjusted
sub-model (3.3) and the original sub-model (2.4). We will see that
the simulation results in Section~5 coincide with these conclusions.

\noindent {\large\bf 4. Calculation for $A$ in the case of large
$r$}

For the convenience of representation, we
here suppose $E(Z)=0, E(U)=0$ and $Cov(Z^\star)=I$.
Lemma A2 given in Appendix shows that the model (3.3) is unbiased if
$A$ is a solution of the following equation:
$$\Sigma_{U,Z^\star}A^{\tau}(AA^{\tau})^{-1}AZ^\star=\Sigma_{U,Z^\star}Z^\star.\eqno(4.1)$$
As was mentioned before, when $r$ is large, a $(r+1)$-dimensional
nonparametric estimation will be involved, which may lead to
inefficient estimation. Thus, we suggest an approximation solution of (4.1), which is a row
vector, that is, $r=1$. Without confusion, we still use the notation $A$ to denote
this row vector. That is, we choose a row vector $A$ such that
$$A^{\tau}AZ^\star=\Sigma_{U,Z^\star}^+\Sigma_{U,Z^\star}Z^\star.\eqno(4.2)$$
By (4.2), an estimator of $A$  can be constructed as follows. Denote
$A=(a_1,\cdots,a_q,a_{q+1})$, $A_k=a_kA$ and
$\Sigma_{U,Z^\star}^+\Sigma_{U,Z^\star}=(D_1^{\tau},\cdots,
D_{q}^{\tau},D_{q+1}^{\tau})^{\tau}$, where $D_k,k=1,\cdots,q+1$,
are $(q+1)$-dimensional row vectors. Then we estimate $A$ via
solving the following optimization problem:
$$\inf\Big\{Q(a_1,\cdots,a_{q+1}):
\sum_{k=1}^{q+1}a_k^2=1\Big\},\eqno(4.3)$$ where
$Q(a_1,\cdots,a_{q+1})=\frac{1}{n}\sum_{i=1}^n\sum_{k=1}^{q+1}\|(A_k-D_k)Z_i^\star\|^2$.
By the Lagrange multiplier, we obtain the estimators of
$A_k,k=1,\cdots,q+1,$ as
$$\hat A_k=\Big(D_k\frac{1}{n}\sum_{i=1}^nZ_i^\star {Z_i^\star}^{\tau}+cc_ke_k/2\Big)\Big(\frac{1}{n}
\sum_{i=1}^nZ_i^\star {Z_i^\star}^{\tau}+c_k
I\Big)^{-1},\eqno(4.4)$$ where $c_k>0$, which is similar to a ridge
parameter, depends on $n$ and tends to zero as $n\rightarrow\infty$,
and $e_k$ is the row vector with $k$-th component being 1 and the
others being zero. Note that the constraint $\|A\|=1$ implies
$\|A_k\|=\pm a_k$. By combining (4.4) with this constraint we get an
estimator of $a_k$ as
$$\hat a_k=\pm \|\hat A_k\|$$ and consequently an estimator of $A$
is obtained by
$$\hat A=(\hat a_1,\cdots,\hat a_q,\hat a_{q+1}).$$

\

\noindent {\large\bf 5. Simulation studies}

In this section we examine the performance of the new method via
simulation studies. By mean squared error (MSE), model prediction
error (PE) and their $std$\,MSE and $std$\,PE as well, we compare
the method with the Gaussian-dantzig selector first. In ultra-high
dimensional scenarios, the Dantzig selector cannot work well, we use
the sure independent screening (SIS) (Fan and Lv 2008) to bring
dimension down to a moderate size and then to make a comparison with
the Gaussian-dantzig selector.  As is well known, there are several
factors that are of great impact on the performance of variable
selection methods: sparse or non-sparse conditions, dimensions $p$
of predictor $X$, correlation structure between the components of
predictor $X$, and variation of the error which can be measured by
theoretical model R-square defined by
$R^2=(Var(Y)-\sigma^2_\varepsilon)/Var(Y)$. Then we will
comprehensively illustrate the theoretical conclusions and performances. 

{\bf Experiment 1.} This experiment is designed mainly for that with
different choices of the theoretical model R-square $R^2$, 
we compare our methods with
Gaussian-dantzig selector. 
In the simulation, to determine the regression coefficients, we
decompose the coefficient vector $\beta$ into two parts: $\beta_I$
and $\beta_{-I},$ where $I$ denotes the set of locations of
significant components of $\beta$. 
Three types of $\beta_I$ are considered:
\\ Type (I): $\beta_I= (1,0.4,0.3,0.5,0.3,0.3,0.3)^{\tau}$
and $I$= \{1,2,3,4,5,6,7\};\\ Type (II): $\beta_I=
(1,0.4,0.3,0.5,0.3,0.3,0.3)^{\tau}$ and $I= \{1,17,33,49,65,81,97\}$;\\
Type (III): $\beta_I= (1,0.4,-0.3,-0.5,0.3,0.3,-0.3)^{\tau}$ and $I=
\{1,2,3,4,5,6,7\}$. \\
To mimic practical scenarios, we set the values of the components
$\beta_{-Ii}$'s of $\beta_{-I}$ as follows. Before performing the
variable selection and estimation, we generate $\beta_{-Ii}$'s from
uniform distribution $\mathcal {U}(-0.5,0.15)$ and the negative
values of them are then set to be zero. Thus the model under study
here is non-sparse. After the coefficient vector $\beta$ is
determined, we consider it as a fixed value vector and regard
$\beta_I$ as the main part of the coefficient vector $\beta$.
We use $\hat I$ to denote the set of subscript of coefficients
$\theta$ in $\beta$, that is the coefficients' subscript of
predictors selected into sub-model. We  assume $X \thicksim
N_{p}(\mu,\Sigma_{X})$, where the components of $\mu$ corresponding
to $I$ are 0 and the others are 2, and the $(i,j)$-th element of
$\Sigma$ satisfies $\Sigma_{ij}=(-\rho)^{\mid i-j\mid}$, $0<\rho<1.$
Furthermore, the error term $\varepsilon$ is assumed to be normally
distributed as $\varepsilon \thicksim N(0,\sigma^2)$. In this
experiment, we choose different $\sigma$ to obtain different type of
full model with different $R^2$. In the simulation procedure, the
kernel function is chosen as the Gaussian kernel
$K(u)=\frac{1}{\sqrt{2\pi}}\exp\{-\frac{u^2}{2}\}$, $A$ is chosen by
(4.4) with $c=2$ and $c_k=0.2$, the choice of parameter $\lambda_p$
in the Dantzig selector is just like that given by Cand\'es and Tao
(2007), which is the empirical maximum of $|X^{\tau}z|_i$ over
several realizations of $z\sim N(0,I_n)$.

The following Tables 1 and 2 report the MSEs and the corresponding
PEs via 200 repetitions. In these tables,  $\hat Y$ is the
prediction via the adjusted model (3.3) that is based on  the full
dataset, $\hat Y_S$ is the prediction via the sub-model (2.4) with
the new estimator $\hat\theta$ defined in (3.7), $\tilde Y_S$ stands
for the prediction via the sub-model (2.4) and the Gaussian-dantzig
selector $\tilde\theta_{S}$. For the definitions of $\hat Y$, $\hat
Y_S$ and $\tilde Y_S$ see (3.8), (3.9) and (3.10), respectively. The
purpose of such a comparison is to see whether the adjustment works
and whether we should use the sub-model (2.4) when the
high-dimensional data are not available (say, too expensive to
collect), whether the new estimator $\hat\theta$ together with the
sub-model (2.4) is helpful for prediction accuracy. The sample size
is $50$, and for the prediction, we perform the experiment with 200
repetitions to compute the proportion $\tau$ of which the prediction
error of $\hat Y_S$ is less than that of $\tilde Y_S$ in the 200
repetitions. The larger $\tau $ is, the better the new prediction
is.

\newpage
\begin{center}
{ \small \centerline{{\bf Table~1.}     MSE, PE and their standard
errors with $n=50,p=100$ and $\rho=0.1$} \tabcolsep0.04in
\vspace{-1ex}
\newsavebox{\tablebox}
\begin{lrbox}{\tablebox}
\begin{tabular}{cc|cc|ccc|c}
  \hline\hline
  & &\multicolumn{2}{c|}  {MSE($std$\,MSE)}  &\multicolumn{3}{c|}{PE($std$\,PE)}& \\
 type&$R^2$&$\hat \theta$   &$\tilde\theta_S$  &$\hat Y$& $\hat Y_S$ & $\tilde Y_S$ &\raisebox{1.5ex}[0pt]{$\tau$}\\\hline
&0.98 &0.0032(0.0118) &0.0866(0.3519) &0.1630(0.0405) &0.2299(0.0535) &1.1587(0.5549) &    200/200 \\
&0.82 & 0.0134(0.0544)&0.1197(0.1654) &0.6603(0.1497) &0.7249(0.1564) &1.4755(0.3475) &    200/200 \\
(I)&0.67 &0.0273(0.1288) &     0.0430(0.1283) & 1.3038(0.2952) &1.3438(0.3018) &1.4821(0.3266) &166/200 \\
&0.50 &     0.0543(0.2387) &0.0694(0.2221) &2.5371(0.5500) &2.5919(0.5633) &2.7176(0.6020) &        142/200 \\
&0.31 &0.1028(0.4689) & 0.1131(0.4876) &4.9199(1.1856) & 4.9960(1.2070) &5.0708(1.1965) &        126/200 \\
\hline
&0.98 &0.0052(0.0202) &0.3540(1.4263) & 0.2584(0.0569) &0.2744(0.0583) &1.1324(2.4262) &        200/200 \\
&0.84 & 0.0162(0.0686) & 0.4087(0.3730) & 0.8310(0.1823) &0.8417(0.1834) &3.7996(0.7909) &        200/200 \\
(II)&0.70 & 0.0292(0.1112) &0.1770 (0.2559) & 1.4761(0.3028) &1.4727(0.3018) & 2.6389(0.5804) &        199/200 \\
&0.53 & 0.0588(0.3024) & 0.0942(0.2988) & 2.8825(0.6534) & 2.8700(0.6460) &3.2707(0.6758) &        171/200 \\
&0.35 & 0.1107(0.6896) & 0.1251(0.6368) & 5.4055 (1.1809) &5.3896(1.1856) &5.6004(1.2280) &        141/200 \\
\hline
&0.98 &0.0028(0.0113) &  0.0879(0.2938) &0.1643(0.0410) &0.2365(0.0537) &1.2282(0.5590) &        200/200 \\
&0.83 & 0.0114(0.0531) &0.0873(0.1589) &0.5874 (0.1332) & 0.6938(0.1533) &1.3483(0.3118) &        200/200 \\
(III)&0.69 & 0.0234(0.0934) &0.1294(0.1667) &1.1922(0.2857) & 1.2445(0.2961) &1.9950(0.4379) &        196/200 \\
&0.51 & 0.0529(0.1715) & 0.0913(0.1775) &  2.6373(0.5788) & 2.7418(0.6098) & 2.9601(0.6288) &        164/200 \\
&0.33 &0.1006(0.5013) & 0.1083(0.5158) & 5.0952(1.2099) &5.1720(1.2241) &5.2372(1.2594) &        119/200 \\
\hline\hline
\end{tabular}
\end{lrbox}
\scalebox{0.9}{\usebox{\tablebox}} }
\end{center}

The simulation results in Table 1 suggest that  the adjustment of
(3.3) works very well, the corresponding estimation ($\hat \theta$) and prediction ($\hat Y$)
are uniformly the best among the  competitors. Further, as we
mentioned, when the full dataset is not available and we thus use
the sub-model of (2.4), the new estimator $\hat \theta$ is also
useful for prediction. It can be seen that $\hat Y_S$ with $\hat \theta$ is better than
$\tilde Y_S$ with the Gaussian-dantzig
selector $\tilde\theta_{S}$, and the value of $\tau$ is larger than 0.7 in 13 cases
out of 15 cases and in
the other 2 cases, it is larger than or about 0.6. 

To provide more information, we also consider the case with higher
correlation between the components of $X$. Table~2 shows that when
$\rho$ is larger, the conclusions about the
comparison are almost identical to those presented in Table~1. 
Thus it concludes that no matter $\rho$ is larger or not, for
different choices of $R^2$, our new method always works quite well.


\newpage

\begin{center}
{ \small \centerline{{\bf Table 2.}     MSE, PE and their standard
errors with $n=50,p=100$ and $\rho=0.7$} \tabcolsep0.04in
\vspace{-1ex}
\begin{lrbox}{\tablebox}
\begin{tabular}{cc|cc|ccc|c}
  \hline\hline
  & &\multicolumn{2}{c|}  {MSE($std$\,MSE)}  &\multicolumn{3}{c|}{PE($std$\,PE)}& \\
 type&$R^2$&$\hat \theta$   &$\tilde\theta_S$  &$\hat Y$& $\hat Y_S$ & $\tilde Y_S$ &\raisebox{1.5ex}[0pt]{$\tau$}\\\hline
&0.96 &0.0136(0.0504) &    0.3285(0.4226) &    0.2472(0.0517) &    0.2706(0.0599) &    1.7397(0.3804) &        200/200 \\
& 0.71 &    0.0253(0.1426) &    0.0709(0.2401) &    0.6530(0.1463) &    0.6945(0.1557) &    1.9892(0.2070) &        197/200 \\
(I)&       0.53 &    0.0373(0.1621) &    0.1108(0.2310) &    1.2779(0.2744) &    1.3235 (0.2861) &    1.5985(0.3736) & 177/200 \\
&       0.35 &    0.0613(0.3122) &    0.0999(0.3289) &    2.3431(0.5342) &    2.3694(0.5395) &    2.6339(0.5799) &        161/200 \\

           &        0.2 &    0.1198(0.6479) &    0.1292(0.6619) &    5.1184(1.2643) &    5.1347(1.2729) &    5.1764(1.2420) &        129/200 \\\hline

           &       0.98 &    0.0122(0.0484) &     0.2730(0.3789) &     0.2648(0.0730) &     0.2809(0.0757) &     1.1952(0.2440)& 200/200 \\

           &       0.84 &    0.0201(0.0924) &    0.1799(0.2037) &    0.6567(0.1453) &    0.6580(0.1452) &    1.6477(0.3560) & 200/200 \\

           (II)&       0.69 & 0.0303(0.1338) &    0.2899(0.4442) & 1.2955(0.2992) &    1.2996(0.3047) &    2.7125(0.5861) & 200/200 \\

           &       0.52 &    0.0644(0.3395) &    0.1141l(0.4388) &    2.5572(0.5558) &    2.5633(0.5582) &    3.2790(0.6834) & 191/200 \\

           &       0.34 &    0.1245(0.5615) &    0.1831(0.6787) &    5.0731(1.1850) &    5.0818(1.1743) &    5.5988(1.2782) & 161/200 \\\hline

           &       0.96 &    0.0239(0.0626) &0.6020(2.1653) &    0.2596(0.0560) &    0.2897(0.0630) &   1.6754(1.4970) &        200/200 \\

           &       0.74 &    0.0315(0.1158) &    0.4401(0.5248) &    0.6435(0.1435) &    0.6485(0.1442) &    2.7859(0.6035) &  200/200 \\

           (III)&       0.56 &    0.0749(0.2373) &    0.1736(0.2679) &    1.3334(0.2947) &    1.4367(0.3217) &    1.8643(0.3965) & 189/200 \\

           &       0.38 &    0.0687(0.3227) &    0.1701(0.3809) &    2.3637(0.4538) &    2.4645(0.4818) &    2.9415(0.5992) & 178/200 \\

           &       0.23 &    0.1740(0.8078) & 0.2446(0.8718) &    4.8488(1.1812) &    4.8887(1.1968) &    5.1471(1.1499) &   145/200 \\

\hline\hline
\end{tabular}
\end{lrbox}
\scalebox{0.9}{\usebox{\tablebox}} }
\end{center}

\

We are now in the position to make another comparison. In
Experiments~2 and 3 below, we do not use the data-driven approach as
given in Experiment 1 to select $\lambda_p$, while manually select
several values to see whether our method works or not. This is
because in the two experiments, it is not our goal to study
shrinkage tuning parameter, but is our goal to see whether the new
method works after we have a sub-model.

{\bf Experiment 2.}  In this experiment, our focus is that with
different choices of the correlation between predictors and
sub-models, we compare our method with others. The distribution of
$X$ is the same as that in Experiment~1 except for
the dimension of the covariate. 
The coefficient vector $\beta_I$ is designed as type (I) above and
$\beta_{-I}$ is designed as in Experiment 1. Thus the model here is
also non-sparse.
 Furthermore, the error term
$\varepsilon$ is assumed to be normally distributed as $\varepsilon
\thicksim N(0,0.2^2)$.  

As different choices of $\lambda_p$ usually lead to different
sub-models, equivalently, to different estimators $\hat I$ of $I$,
we then consider different choices of $\lambda_p$ in the simulation
study. 
The setting is as follows. For $n=50,p=100$ and
$\rho=0.1,0.3,0.5,0.7,$ we consider two cases
for each $\rho$: \\
$\rho=0.1:\\
$ Case 1. $\lambda_p=3.97,$ $I$=\{1,2,3,4,5,6,7\}, $\hat I$=\{1,     2,     3,     4,     5,     6,     7 \}\\
Case 2. $\lambda_p=6.53,$ $I$=\{1,2,3,4,5,6,7\}, $\hat I$=\{1,     3,     4,     6,    95 \}\\
$\rho=0.3:\\$ Case 1. $\lambda_p=3.32,$ $I$=\{1,2,3,4,5,6,7\}, $\hat I$=\{1,     2,     3,     4,     5,     6 \}\\
Case 2. $\lambda_p=6.77,$ $I$=\{1,2,3,4,5,6,7\}, $\hat I$=\{ 1,     2,     4,     6,    23 \}\\
$\rho=0.5:\\$ Case 1. $\lambda_p=3.72,$ $I$=\{1,2,3,4,5,6,7\}, $\hat I$=\{1,     2,     4,     5,     6,     7 \}\\
Case 2. $\lambda_p=7.29,$ $I$=\{1,2,3,4,5,6,7\}, $\hat I$=\{1,     4,     5,     7,    41,    58,    72 \}\\
$\rho=0.7:\\$ Case 1. $\lambda_p=3.50,$ $I$=\{1,2,3,4,5,6,7\}, $\hat I$=\{1,     3,     4,     7,    41,    75\}\\
Case 2. $\lambda_p=7.22,$ $I$=\{1,2,3,4,5,6,7\}, $\hat I$=\{1,     4,     7,    51,    64,    67,    68,    83 \}\\

\begin{center}
{ \small \centerline{{\bf Table 3.}     MSE, PE and their standard
errors with $n=50,p=100,S=7$ } \tabcolsep0.045in \vspace{-1ex}
\begin{lrbox}{\tablebox}
\begin{tabular}{cc|cc|ccc|c}
  \hline\hline
  & &\multicolumn{2}{c|}  {MSE($std$\,MSE)}  &\multicolumn{3}{c|}{PE($std$\,PE)}& \\
$\rho$&Case&$\hat \theta$   &$\tilde\theta_S$  &$\hat Y$& $\hat Y_S$&$\tilde Y_S$ &\raisebox{1.5ex}[0pt]{$\tau$}\\\hline
                          &1&   0.0052(0.0242) &    0.2929(0.3877) &   0.2580(0.0528) &   0.2612(0.0527) &   3.0195(0.6691) &  200/200\\
\raisebox{1.5ex}[0pt]{0.1}&2 &   0.0104(0.0357) &   0.2347(0.1784) &   0.5135(0.1074) &   0.6430(0.1282) &    5.921(0.4172) &200 /200 \\\hline
                          &1 &   0.0070(0.0289) &   0.4067(1.6692) &   0.2732(0.0590) &   0.3324(0.0735) &   5.6406(1.8289) &200/200\\
\raisebox{1.5ex}[0pt]{0.3}&2 &     0.0163(0.0458) &   0.5048(0.4107) &   0.4048(0.0881) &   0.5014(0.1078) &   6.4471(0.7697) &200/200\\\hline
                          &1 &   0.0079(0.0336) &   0.4826(1.9425) &    0.2436(0.0551) &    0.3053(0.0674) &   5.8204(1.8152) & 200/200\\
\raisebox{1.5ex}[0pt]{0.5}&2 &   0.0136(0.0512) &    0.1532(0.1835) &    0.3655(0.0841) &    0.4245(0.0914) &    6.4357(0.3262) &200/200\\\hline
                          &1 &   0.0157(0.0602) &    0.2296(0.2970) &    0.2688(0.0580) &    0.3198(0.0711) &    6.6313(0.3560) &200/200\\
\raisebox{1.5ex}[0pt]{0.7}&2 &   0.0149(0.0637) &    0.1914(0.1420) &    0.2974(0.0624) &    0.3225(0.0672) &    7.5435(0.1169) &197/200\\
\hline\hline
 \end{tabular}

\end{lrbox}
 \scalebox{0.9}{\usebox{\tablebox}} }
\end{center}

From Table~3, we can see clearly that the correlation is of impact
on the performance of the variable selection methods:  the
estimation gets worse with larger $\rho$. However, the new method
uniformly works much better than the Gaussian Dantzig selector, when
we compare the performance of the methods with different values of
$\lambda_p$ and then with different sub-models. We can see that in
case I, the sub-models are more accurate than those in case II in
the sense that they can contain more significant predictors we want
to select. Then, the estimation based on the Gaussian Dantzig
selector can work better and so can the new method. 

In the following, we consider data with higher-dimension.

{\bf Experiment 3.} In this experiment $\beta_{-I}$ is designed as
in Experiment 1. Thus the model here is also non-sparse. For very
large $p$, the Dantzig selector method alone cannot work well. Thus,
we use the sure independent screening (SIS,  Fan and Lv 2008) to
reduce the number of predictors to a moderate scale that is below
the sample size, and then perform the variable selection and
parameter estimation afterwards by the
Gaussian Dantzig selector and our adjustment method. 
The experiment conditions are designed as: $$\beta_I=(1.0, -1.5,
2.0, 1.1, -3.0, 1.2, 1.8, -2.5, -2.0, 1.0)^{\tau},n=100,p=1000;$$
\noindent $\rho$=0.1:  \\
Case 1. $\lambda_p$=4.50,  $I$=\{1,2,3,4,5,6,7,8,9,10\}, $\hat I=\{1,     3  ,   5 ,    6 ,    7  ,   8,     9 ,  318,   514 ,  723,   760
\}$;\\
Case 2. $\lambda_p$=7.30,  $I$=\{1,2,3,4,5,6,7,8,9,10\}, $\hat I=\{ 2 ,    3 ,    5,     8,   515 ,  886
 \}$.\\
$\rho$=0.5: \\
Case 1. $\lambda_p$=3.56,  $I$=\{1,2,3,4,5,6,7,8,9,10\}, $\hat I=\{1  ,   2  ,   5 ,    7,     8 ,    9 ,  846 ,  878 ,  976
  \}$;\\
Case 2. $\lambda_p$=6.92,  $I$=\{1,2,3,4,5,6,7,8,9,10\}, $\hat I=\{
2   ,  3 ,    5 ,    8 ,   10,   882 ,  963
 \}$.\\
$\rho$=0.9:\\
Case 1. $\lambda_p$=1.80,  $I$=\{1,2,3,4,5,6,7,8,9,10\}, $\hat I=\{  3,     5,     8,    10 ,  415 ,  432
 \}$;\\
Case 2. $\lambda_p$=5.83,  $I$=\{1,2,3,4,5,6,7,8,9,10\}, $\hat I=\{
2 ,   3   ,  5 ,  114 ,  121 ,  839 ,  853,   882 ,  984\}$.

With this design, the $\lambda_p$ in case 1 results in that  more
significant predictors are selected into the sub-model than those in
case 2 so that we can see the performance of the adjustment method.

\newpage
\begin{center}
{ \small \centerline{{\bf Table 4.}     MSE, PE and their standard
errors with $n=100$ and $p=1000$ } \tabcolsep0.04in \vspace{-1ex}
\begin{lrbox}{\tablebox}
\begin{tabular}{cc|cc|ccc|c}
  \hline\hline
  & &\multicolumn{2}{c|}  {MSE($std$\,MSE)}  &\multicolumn{3}{c|}{PE($std$\,PE)}& \\
$\rho$&Case&$\hat \theta$   &$\tilde\theta_S$  &$\hat Y$& $\hat
Y_S$&$\tilde Y_S$ &\raisebox{1.5ex}[0pt]{$\tau$}\\\hline
                          &1&  0.7588(0.3497) &    71.4031(7.5501) &     6.8104(1.5485) &     8.0107(1.6574) &    94.7515(19.2968) & 200/200 \\
\raisebox{1.5ex}[0pt]{0.1}&2&  0.8523(0.5343) &   122.8426(15.0952)
&    13.1274(2.7772) &    16.0812(3.4160) &   189.7134(34.8081) &
200/200 \\\hline
                          &1&  3.6170(1.1823) &   104.8420(13.5089) &     9.9151(1.9902) &    11.2352(2.2316) &   133.4762(26.5058) &200/200\\
\raisebox{1.5ex}[0pt]{0.5}&2&3.4771(1.2683) &    92.3485(12.5122) &
11.6643(2.6704) &    12.7811(2.8941) &   134.3821(24.4896)
&200/200\\\hline
                          &1&  5.9027(2.7039) &   107.6118(23.4383) &     8.2842(1.6181) &    11.3518(2.1745) &   148.3143(27.4828) & 200/200\\
\raisebox{1.5ex}[0pt]{0.9}&2& 3.8963(2.1760) &    59.1525(11.3152) &    10.8033(2.1411) &    12.9395(2.4835) & 68.7272(13.4061) & 200/200\\
\hline\hline
 \end{tabular}

\end{lrbox}
\scalebox{0.85}{\usebox{\tablebox}} }
\end{center}

\

From Table 4, we have the conclusion that the SIS does work to
reduce the dimension so that the Gaussian Dantzig selector and our
method can be performed. Whether the correlation coefficient is
small or large (the values of $\rho$ change from 0.1 to 0.9), the
new method works better than the Gaussian Dantzig selector. The
conclusions are almost identical to those when $p$ is much smaller
in Experiments 1 and 2. Thus, we do not give more comments here.
Further, by comparing the results of case 1 and case 2, we can see
that the adjustment can work better when the sub-model is not well
selected. 

 In the following we further check the effect of model
size when the dimension is larger. In doing so, we choose $n=150,
p=2000$, $\rho=0.3$; \\
For $\beta_I=(4.0, -1.5, 6.0, -2.1,
-3.0)^{\tau}$, consider two cases:  \\
Case 1. $\lambda_p$=3.45, $I$=\{1,2,3,4,5\}, $\hat I=\{
1,2,3,4,5,15,1099,1733
 \}$;\\
Case 2. $\lambda_p$=8.36, $I$=\{1,2,3,4,5\}, $\hat I=\{1,3,554,908
  \}$.\\
For $\beta_I=(4.0, -1.5, 6.0, -2.1, -3.0, 1.2, 3.8, -2.5, -2.0, 7.0)^{\tau}$, consider two cases:\\
Case 1. $\lambda_p$=3.02, $I$=\{1,2,3,4,5,6,7,8,9,10\}, $\hat I=\{1,2,3,5,7,8,9,10,1701
 \}$;\\
Case 2. $\lambda_p$=9.08, $I$=\{1,2,3,4,5,6,7,8,9,10\}, $\hat I=\{1,3,5,7,8   \}$.\\

\newpage
\begin{center}
{ \small \centerline{{\bf Table 5.}     MSE, PE and their standard
errors with $n=150,p=2000,\rho=0.3$ } \tabcolsep0.02in \vspace{-1ex}
\begin{lrbox}{\tablebox}
\begin{tabular}{cc|cc|ccc|c}
  \hline\hline
  & &\multicolumn{2}{c|}  {MSE($std$\,MSE)}  &\multicolumn{3}{c|}{PE($std$\,PE)}& \\
$$S$$&Case&$\hat \theta$   &$\tilde\theta_S$  &$\hat Y$& $\hat
Y_S$&$\tilde Y_S$ &\raisebox{1.5ex}[0pt]{$\tau$}\\\hline
                          &1&  0.4245(0.2102) &262.6392(21.2109) & 6.4015(1.3038) &     6.3439(1.2879) &322.9945(62.6228) &200/200 \\
\raisebox{1.5ex}[0pt]{5}&2&  1.9510(1.0923) & 359.5838(32.4150) & 24.1959(4.8932) &    24.8013(5.1629) & 559.3584(98.1216) &200/200 \\\hline
                          &1&  0.8799(0.5108) & 498.7862(59.0383) &10.6009(2.3903) &    12.3505(2.6381) &946.3400(175.1009) & 200/200\\
\raisebox{1.5ex}[0pt]{10}&2&  1.8524(0.7599) &68.1862(43.3612) &15.0471(2.8069) & 16.9161 (3.1755) &  1623.4936(111.5972) & 200/200\\
\hline\hline
 \end{tabular}

\end{lrbox}
\scalebox{0.9}{\usebox{\tablebox}} }
\end{center}

The results in Table 5 show that the SIS is again useful for
reducing the dimension for the use of the Gaussian Dantzig selector
and our method, and furthermore the new method works better than the
Gaussian Dantzig selector. On the other hand, when the number of
significant predictors is smaller, estimation accuracy can be better
with smaller MSE and PE. In other words, when the number of
significant predictors is smaller, variable selection can perform
better and sub-model can be
more accurate (case 1 with 5 significant predictors). 


{\bf Experiment 4.} This experiment is designed for checking that
although our method is designed for the non-sparse model, it is also comparable to the method designed for sparse model when the true
model is sparse indeed. We also consider three type of $\beta$ which
is the same as those in Experiment 1 except that all components of
$\beta_{-I}$ are zero. The simulation result is reported in Table~6
below.

\newpage

\begin{center}
{ \small \centerline{{\bf Table 6.}     MSE, PE and their standard
errors with $n=50$ and $p=100$ for the sparse case} \tabcolsep0.04in
\vspace{-1ex}
\begin{lrbox}{\tablebox}
\begin{tabular}{cc|cc|ccc|c}
  \hline\hline
  & &\multicolumn{2}{c|}  {MSE($std$\,MSE)}  &\multicolumn{3}{c|}{PE($std$\,PE)}& \\
 type&$\rho$&$\hat \theta$   &$\tilde\theta_S$  &$\hat Y$& $\hat Y_S$ & $\tilde Y_S$ &\raisebox{1.5ex}[0pt]{$\tau$}\\\hline
   &0.1 &0.9938$\times 10^{-3}$(0.0040) &   0.9324$\times 10^{-3}$(0.0037) &    0.0485(0.0114) &    0.0481(0.0113) &    0.0469(0.0109) &        71/200 \\
   & 0.3 &    0.0013(0.0051) &    0.0033(0.0118) &
    0.0668        (0.0152) &    0.1373(0.0262 ) &    0.1440(0.0290) &      134/200 \\
(I)&       0.5 &
    0.0036    (0.0128) &    0.0068(0.0239) &    0.1856(0.0429)  &  0.2905(0.0603) &   0.2999(0.0640) & 138/200 \\
   &       0.7 &    0.0066(0.0187) &   0.0100(0.0278) &    0.2485(0.0578)  &  0.3288(0.0713) &   0.3311(0.0708) &        115/200 \\

   &        0.9 &    0.1198(0.6479) &    0.1292(0.6619) &    0.3506(0.0758)&    0.4624(0.0881)   & 0.4630(0.0867) &        99/200 \\\hline

    &       0.1 &    0.0010(0.0039)  &  0.0010(0.0039)  &  0.0482(0.0112) &   0.0479(0.0109)  &  0.0468(0.0102)
& 73/200 \\

    &       0.3 &     0.0028(0.0105)   & 0.0029(0.0110) & 0.1473(0.0315)   & 0.1529

(0.0324)  &  0.1485(0.0330) & 80/200 \\

(II)&       0.5 &  0.0029(0.0104) &   0.0030(0.0113)&
    0.1462(0.0315)    &0.1526(0.0328)  &  0.1496(0.0329) & 85/200 \\
    &       0.7 &     0.0052(0.0160)   & 0.0072(0.0209) &   0.2832(0.0626) &   0.3460(0.0736)  &  0.3477(0.0743)& 114/200 \\

    &       0.9 &   0.0059(0.0169) &   0.0220(0.0460) &   0.3360(0.0921)   & 0.5392(0.1333)   & 0.5250(0.1202)
 & 83/200 \\\hline

           &       0.1 &    0.9773$\times 10^{-3}$(0.0040)  &  0.9425$\times 10^{-3}$(0.0039) &    0.0483(0.0108) &   0.0479(0.0108) &   0.0468(0.0105) &        71/200 \\

           &       0.3 &    0.0034(0.0120)  &  0.0060(0.0218)&    0.1697(0.0383)  &  0.2547(0.0516)  &  0.2629(0.0547) &  132/200 \\

    (III)&       0.5 &    0.0046(0.0142)  &  0.0073(0.0215)&    0.2386(0.0573) &   0.3260

(0.0700)  &  0.3269(0.0679) & 122/200 \\

           &       0.7 &    0.0076(0.0200)   & 0.0114(0.0277) & 0.3633(0.0775)  &  0.4997

(0.1016)  &  0.5030(0.1050) & 112/200 \\

           &       0.9 &    0.0100(0.0229)  &  0.0148(0.0321)&   0.4641(0.1067)&    0.6082

(0.1264)  &  0.6012(0.1224) &   83/200 \\

\hline\hline
\end{tabular}
\end{lrbox}
\scalebox{0.9}{\usebox{\tablebox}} }
\end{center}

\

From this table, we can see that even in sparse cases, for every
type of $\beta$, the new estimator $\hat \theta$ is in almost all
cases better than   $\tilde\theta_S$ is in the sense of smaller MSE.
This is also the case for prediction: $\hat Y$ has smaller
prediction error than $\tilde Y_S$ does when $\rho\ge 0.1$. It is
not surprise that $\hat Y_S$ cannot be as good as its performance in
non-sparse cases, but still comparable to $\tilde Y_S$. From the
table, we can see that $\tilde Y_S$ is usually better than $\hat
Y_S$ when $\rho$ is either $0.1$ or $0.9$ and $\tau<0.5$ whereas
when $0.3\le \rho \le 0.7$, the prediction error of $\tilde Y_S$ is
larger and $\tau> 0.5$ except for the cases with $\rho = 0.3, \,
0.7$ in type II of $\beta$. Overall, the new method is still
comparable to the classical method in the sparse models under study.

In summary, the results in the six tables above obviously show the
superiority of the new estimator $\hat\theta$ and the new sub-model
(3.3)/the sub-model (2.4) over the others in the sense with smaller
MSEs, PEs and standard errors, and large proportion $\tau$ in non-sparse models.  The good
performance holds for different combinations of the sizes of
selected sub-models (values of $\lambda_p$), $n,p,S,I$, $R^2$ and
the correlation between the components of $X$. The new method is
particularly useful when a submodel, as a working model, is very
different from underlying true model. Thus,  the adjustment method
is  worth of recommendation. Also it is comparable to the classical method in sparse case, suggesting its robustness against model structure. However, as a trade-off, the
adjustment method involves nonparametric estimation, although
low-dimensional ones. It makes estimation not as simple as that obtained by the existing ones. Thus, we may consider using it after a
check whether the submodel is significantly biased. The relevant
research is ongoing.

\

\noindent{\large\bf  Supplementary Materials.}
\begin{description}

     \item[Proofs of the theorems:] The pdf file  ``supplement-1.pdf" containing detailed proofs of the lemmas and theorems.
     \item[Matlab package for DANTZIG CODE routine:] Matlab package "DANTZIG CODE"  containing the codes. (WinRAR file)

\end{description}


\bibliography{dantzig_two_stage}
\

\leftline{\large\bf References}

\begin{description}

\baselineskip=15pt

\item Bai, Z. and Saranadasa, H. (1996). Effect of high dimension:
by an example of a two sample problem. {\it Statistica Sinica}, {\bf
6}, 311-329.

\item Bickel, P.J., Ritov, Y. and Tsybakov, A. B. (2009). Simultaneous
analysis of lasso and Dantzig selector. {\it Ann.  Statist.}, {\bf
37}, 1705-1732.

\item Cai, T., Zhang, C. and Zhou, H. H. (2010). Optimal rates of
convergence for covariance matrix estimation. {\it Ann. Statist.},
{\bf 38}, 2118-2144.

\item Cand\'es, E. J. and Tao, T. (2005). Decoding by linear
programming. {\it IEEE Trans. Inform. Theory} {\bf 51}, 4203-4215.

\item Cand\'es, E. J. and Tao, T. (2006). Near-optimal signal recovery
from random projections: Universal encoding strategies? {\it IEEE
Trans. Inform. Theory} {\bf 52}, 5406-5425.

\item Cand\'es, E. J. and Tao, T. (2007). The Dantzig selector:
statistical estimation when $p$ is much larger than $n$. {\it Ann.
Statist.} {\bf 35}, 2313-2351.

\item Chen, S. X. and Qin, Y. L. (2009). A two sample test for high
dimensional data with applications to gene-set testing. {\it Ann.
Statist.} (to appear).

\item Diaconis, P. and Freedman, D. (1984). Asymptotics of graphical
projection pursuit. {\it Ann. Statist.}, {\bf 12}, 793-815.

\item Fan, J. and Li, R. (2001). Variable selection via nonconcave
penalized likelihood and its oracle properties. {\it J. Am. Statist.
Ass.}, {\bf 96}.

\item Fan, J. and Peng, H. (2004). Nonconcave penalized likelihood
with a diverging number of parameters. {\it Ann. Statist.}, {\bf
32}, 928-961.

\item Fan, J., Peng, H. and Huang, T. (2005). Semilinear
high-dimensional model for normalized of microarray data: a
theoretical analysis and partial consistency. {\it J. Am. Statist.
Ass.}, (with discussion), {\bf 100}, 781-813.

\item Fan, J. and Lv, J. (2008). Sure independence screening for
ultrahigh dimensional feature space. {\it J. R. Statist. Soc.} B
{\bf 70}, 849-911.

\item H\"ardle, W., Liang, H. and Gao, T. (2000). {\it Partially linear
models.} Physica Verlag.

\item
Hall, P. and Li, K. C. (1993). On almost linearity of low
dimensional projection from high dimensional data. {\it Ann.
Statist.} {\bf 21}, 867-889.


\item Huang, J., Horowitz, J. L. and Ma, S. (2008). Asymptotic
properties of Bridge estimators in sparse high-dimensional
regression models. {\it Ann. Statist.}, {\bf 36}, 2587-2613.

\item Huber, P. J. (1973). Robust regression: Asymptotic,
conjectures  and Montes Carlo. {\it Ann. Statist.}, {\bf 1},
799-821.

\item James, G. M. and Radchenko, P. (2009). A generalized Dantzig
selector with shrinkage tuning. {\it Biometrika}, {\bf 96}, 323-337.

\item James, G. M. , Radchenko, P. and Lv, J. C. (2009).
Dasso: connections between the Dantzig selector and lasso. {\it
Journal of the Royal Statistical Society, Series B}, {\bf 71},
127-142.



\item Kosorok, M. and Ma, S. (2007). Marginal asymptotics for the
large $p$, small $n$ paradigm: With application to Microarray data.
{\it Ann. statist.}, {\bf 35}, 1456-1486.

\item Kuelbs, J. and Anand, N. W. (2009). Asymptotic inference for
high dimensional data. {\it Ann. Statist.} (to appear) .

\item Lam, C. and Fan, J. (2008). Profile-kernel likelihood inference
with diverging number of parameters. {\it Ann. Statist.}, {\bf 36},
2232-2260.

%

\item Li, G. R., Zhu, L. X. and Lin, L. (2008). Empirical Likelihood for a
varying coefficient partially linear model with diverging number of
parameters. {\it Manuscript}.

\item Malgouyres, F. and Zeng, T. (2009). A predual proximal point algorithm solving a non negative basis pursuit denoising model. {\it Int. J. comput Vis.}, {\bf 83}, 294-311.

%

\item Portnoy, S. (1988). Asymptotic behavior of likelihood methods
for exponential families when the number of parameters tends to
infinity. {\it Ann. Statist.}, {\bf 16}, 356-366.
%

\item Rao, C. R. and Mitra, S. K. (1971). {\it Generalized inverse of
matrices and its application.} John Wily.

\item
Robins, J. and Vaart, A. V. D. (2006). Adaptive nonparametric
confidence sets. {\it Ann. Statist.}, {\bf 34}, 229-253.



\item Zhang, C. H. (2010),
Nearly unbiased variable selection under minimax concave penalty.
\textit{The Annals of statistics}, {\bf 38}, 894--942.

\item Zhang, C. and Huang, J. (2008). The sparsity and bias of the
LASSO selection in high-dimensional linear regression. {\it Ann.
Statist.}, {\bf 36}, 1567-1594.

\item Zhao, P. and and Yu, B. (2006). On model selection consistency
of Lasso. {\it Journal of Machine Learning Research}, {\bf 7},
2541-2563.

\item Zou, H. (2006). The adaptive lasso and its oracle properties. {\it J. Am. Statist.
Assoc.}, {\bf 101}, 1418-1429.

\end{description}

\end{document}